\documentclass{aa}  
\usepackage{txfonts}
\usepackage{epic}
\usepackage{eepic}
\usepackage{graphicx}
\usepackage{epstopdf}
\usepackage{amsmath}
\usepackage{subfig}
\usepackage{xcolor}
\usepackage[]{natbib}
\usepackage{todonotes}
\usepackage{mathtools}

\def\br{\mbox{\boldmath$r$}}
\def\be{\mbox{\boldmath$e$}}
\def\bxi{\boldsymbol \xi}
\def\ii{\textrm{i}}
\def\Cref{\overline{C}^{\ \textrm{ref}}}
\defcitealias{1993AA...268..309T}{TG93}
\defcitealias{1996AA...305L..33Z}{ZSB96}

\begin{document} 
\title{Modelling continuum intensity perturbations caused by solar~acoustic~oscillations}
 \titlerunning{Modelling continuum intensity perturbations}
   \author{N.M.~Kostogryz \inst{1}, D.~Fournier \inst{1}
          \and
          L. Gizon\inst{1,2,3}
          }

   \institute{\inst{1} Max-Planck-Institut f\"ur Sonnensystemforschung, Justus-von-Liebig-Weg 3, 37077 G\"ottingen, Germany\\ \email{kostogryz@mps.mpg.de}\\ 
   \inst{2} Institut f\"ur Astrophysik, Georg-August-Universit\"at G\"ottingen, Friedrich-Hund-Platz 1, 37077 G\"ottingen, Germany\\
   \inst{3}Center for Space Science, NYUAD Institute, New York University Abu Dhabi, PO Box 129188, Abu Dhabi, UAE}
             
\date{}

\abstract
  {
  Helioseismology is the study of the Sun's interior using  observations of  oscillations at the surface. It suffers from systematic errors, such as a center-to-limb error in travel-time measurements. Understanding these errors requires a good understanding of the nontrivial relationship between wave displacement and helioseismic observables (intensity or velocity).
  }
  {
  The wave displacement causes perturbations in the atmospheric thermodynamical quantities which, in turn, perturb the opacity, the optical depth, the source function, and the local ray geometry, thus affecting the emergent intensity. 
  We aim to establish the most complete relationship up to now between the wave displacement and the emergent intensity perturbation by solving the radiative transfer problem in the perturbed atmosphere. 
  }
  {
  We derive an expression for the emergent intensity perturbation caused by acoustic oscillations at any point on the solar disk by applying first-order perturbation theory. As input perturbations, we consider adiabatic modes of oscillation of different degrees in a spherically-symmetric solar model. The background and the perturbed intensities are computed by solving the radiative transfer equation considering the main sources of opacity in the continuum (absorption and scattering).
  }  
   { 
    We find that, for all modes, the perturbations to the thermodynamical quantities are not sufficient to model the intensity perturbations: the geometrical effects due to the wave displacement must always be taken into account as they lead to a difference in amplitude and a phase shift between temperature perturbations at the surface and emergent intensity perturbations. The closer to the limb, the larger the differences.
    For modes with eigenfrequencies around $3~\rm{mHz}$, we found that the radial and horizontal components of the wave displacement are important in particular for high-degree modes.
   }
   {
   This work presents improvements for the computation of the intensity perturbations, in particular for high-degree modes, and explains differences in intensity computations in earlier works. The phase shifts and amplitude differences between the temperature and intensity perturbations increase towards the limb. This should help to interpret some of the systematic center-to-limb effects observed in local helioseismology. The computations are fast (3~s for 2000 positions and one frequency for one core) and can be parallelized. This work can be extended to model the line-of-sight velocity observable. 
   }


\maketitle


\newpage

\section{Introduction} \label{Intro} 
Local helioseismology aims at probing the subsurface structure and the dynamics of the solar convection zone. There are a variety of helioseismology techniques, such as ring diagram analysis, time-distance analysis, and helioseismic holography \cite[see, e.g. review by][]{2005LRSP....2....6G}. All of these techniques suffer from substantial and unexplained systematic effects. 
One of such effect, a systematic center-to-limb variation, was shown in helioseismic travel-time measurements by \cite{2012ApJ...749L...5Z}. They applied time-distance analysis to different observables from the Helioseismic and Magnetic Imager  (\citealp[HMI,][]{2012SoPh..275..207S}) instrument: 
continuum intensity, line-core, and line-depth intensities, and Doppler velocity. For each observable, they observed strong variations of the travel times in the East-West direction as a function of longitude which cannot be caused by any physical flow. This center-to-limb effect in travel time manifests differently in different HMI observables, e.g. it is significantly larger in continuum intensity ($\sim 10$~s) than in the line-core intensity and Doppler velocity ($\sim 2$~s). Moreover, some observables, i.e. line-core and line-depth intensities show opposite trends.
As this effect is not understood, \cite{2012ApJ...749L...5Z} proposed
to apply a simple correction to the North-South travel-time differences 
by subtracting the component of the East-West travel-time differences that is antisymmetric across the central meridian. 
This simple procedure has been used to infer the meridional flow from the corrected North-South travel times \citep{2012ApJ...749L...5Z, 2020Sci...368.1469G}.
In order to understand whether this procedure is valid, it is important to understand the physical and/or instrumental reasons for this center-to-limb effect. 
In the present paper we focus on the  geometrical and radiative transfer effects that may affect the continuum intensity. \cite{2018A&A...619A..99L} and \cite{2020Sci...368.1469G} noted however that the center-to-limb effect seen in HMI travel times varies strongly with time over the course of the mission; thus an instrumental component (which we do not address here) is expected as well.

A better understanding of the physical reason for the center-to-limb effect requires determining the relationship between solar oscillations and helioseismic observables. In helioseismology, observables are often assumed to be directly proportional to temperature fluctuations or to the line-of-sight projection of the wave displacement at fixed radius. A step forward in understanding the center-to-limb effect is to compute the wave perturbations at different heights in the photosphere where the maximum of solar intensity forms and which depends on position on the solar disk. 
\cite{2013SoPh..287..129W} proposed that the phase of the p-mode eigenfunctions combined with the dependence of formation height of solar intensity with heliocentric angle may lead to a center-to-limb effect in helioseismic observables. \cite{2012ApJ...760L...1B} proposed that another contribution may be  due to the interaction of p-modes with granulation, viewed from different lines of sight. However they stress that a full quantitative prediction of the center-to-limb effect requires solving the radiative transfer problem in the atmosphere perturbed by p-modes.

\begin{table*}
    \caption{Computation of intensity perturbation by different authors.}
    \centering
    \begin{tabular}{c|c|c|c|c|}
        Publication & Radiative transfer & Opacity & $\xi_r$ & $\xi_h$ \\
        \hline
        \citet{1990AA...227..563B} & Eddington approx.$^*$ & no & yes & no \\
        \citet{1993AA...268..309T} & yes & only $\rm{H^-}$ & yes & no \\
        \cite{1995ASPC...76..338S} & yes & yes  & no  & no \\
        \cite{1996AA...305L..33Z} & yes & yes  & no  &  no\\
        This work & yes & yes & yes & yes \\
    \end{tabular}
    \tablefoot{$^*$ The Eddington approximation was used to derive the mean intensity but the intensity was not calculated at any points on the disk. The columns $\xi_r$ (resp. $\xi_h$) means that the radial (resp. horizontal) part of the surface perturbation is taken into account (Sect.~\ref{Displacement}).  }
    \label{tab:list}
\end{table*}

Various approximations have been considered to compute the disk-integrated intensity perturbations caused by acoustic and gravity modes.
The pioneering study from \citet{1977AcA....27..203D} derived the expression of emergent flux perturbation and surface distortion assuming that the emergent intensity perturbation caused by the oscillations is known. Further developments have been done to consider non-adiabatic and non-radial oscillations \citep[e.g.][]{1988ESASP.286..387P,1990AA...227..563B} but the emergent intensity was computed using the  black-body or Eddington approximation so neglecting the perturbations of the opacity induced by the oscillations. The assumption that the brightness fluctuation have the same phase and amplitudes as temperature perturbation was commonly used. However, further studies showed that this approximation is not correct. An important improvement in this direction was done by \citet[][hereafter \citetalias{1993AA...268..309T}]{1993AA...268..309T} who derived a more complete expression for emergent intensity taking into account opacity perturbations caused by solar oscillations of low-degree modes in a non-grey atmosphere. It was shown that the emergent intensity fluctuations are proportional not only to temperature but also to density perturbations, and both contributions are equally important. However, opacity was computed only with the bound-free transitions of $\rm {H^-}$ which is the main source of opacity in the visible wavelength range but not the only one.

Later \citet{1995ASPC...76..338S} and  \citet[][hereafter \citetalias{1996AA...305L..33Z}]{1996AA...305L..33Z}  took into account various sources of opacity in the continuum, but neglected the geometrical term due to a compression or an expansion of the atmospheric layers due to the solar oscillations. They obtained a slightly different emergent intensity than \citetalias{1993AA...268..309T} and explained this difference by the sources of opacity neglected by  \citetalias{1993AA...268..309T}. We will discuss further this point in Sect.~\ref{sect:low}.
Table~\ref{tab:list} shows a summary of the main differences between previous  studies and our work. 

All the previous efforts concerning the computation of emergent intensity perturbations were done for the oscillations of the modes with harmonic degree $0 \le l \le 2$, as only these modes are visible in integrated light. However, the techniques of helioseismology applied to resolved images of the Sun take all modes into consideration, i.e. from pure radial ($l=0$) mode for which the horizontal component of wave displacement ($\xi_h$) is zero, up to $l=1500$ modes for which $\xi_h$ and $\xi_r$ are both important at frequency $3\ \rm{mHz}$. Up to now, $\xi_h$ was not considered at all in any of the previous studies (Table.~\ref{tab:list}). In this paper, we establish the connection between the observables, i.e. continuum intensity, and oscillations by solving the radiative transfer equation in the perturbed solar atmosphere. We take into account that the perturbation is caused by both the radial and the horizontal components of the wave displacement vector. This study will be extended to the modelling of the spectral line and Doppler velocity and finally to the interpretation of the travel-time measurements in future work.

The structure of this paper is as follows. In the next section, we derive the expression for emergent intensity perturbations induced by oscillations of different modes taking into account the radial and horizontal components of the wave displacement. Section~\ref{sect:num} describes the numerical methods used to compute the adiabatic oscillations and the opacity in the atmosphere. Section~\ref{results} validates numerically the theoretical derivation of emergent intensity perturbations and presents the results for the intensity fluctuations due to p-modes oscillations. 
Finally, we summarize our study and discuss possible extensions.

\section{Intensity perturbation} \label{s:IntPert}

\subsection{Coordinate systems}
\begin{figure}
    \centering
     \includegraphics[trim={3cm 12cm 12cm 0cm},clip, width=0.9\linewidth]{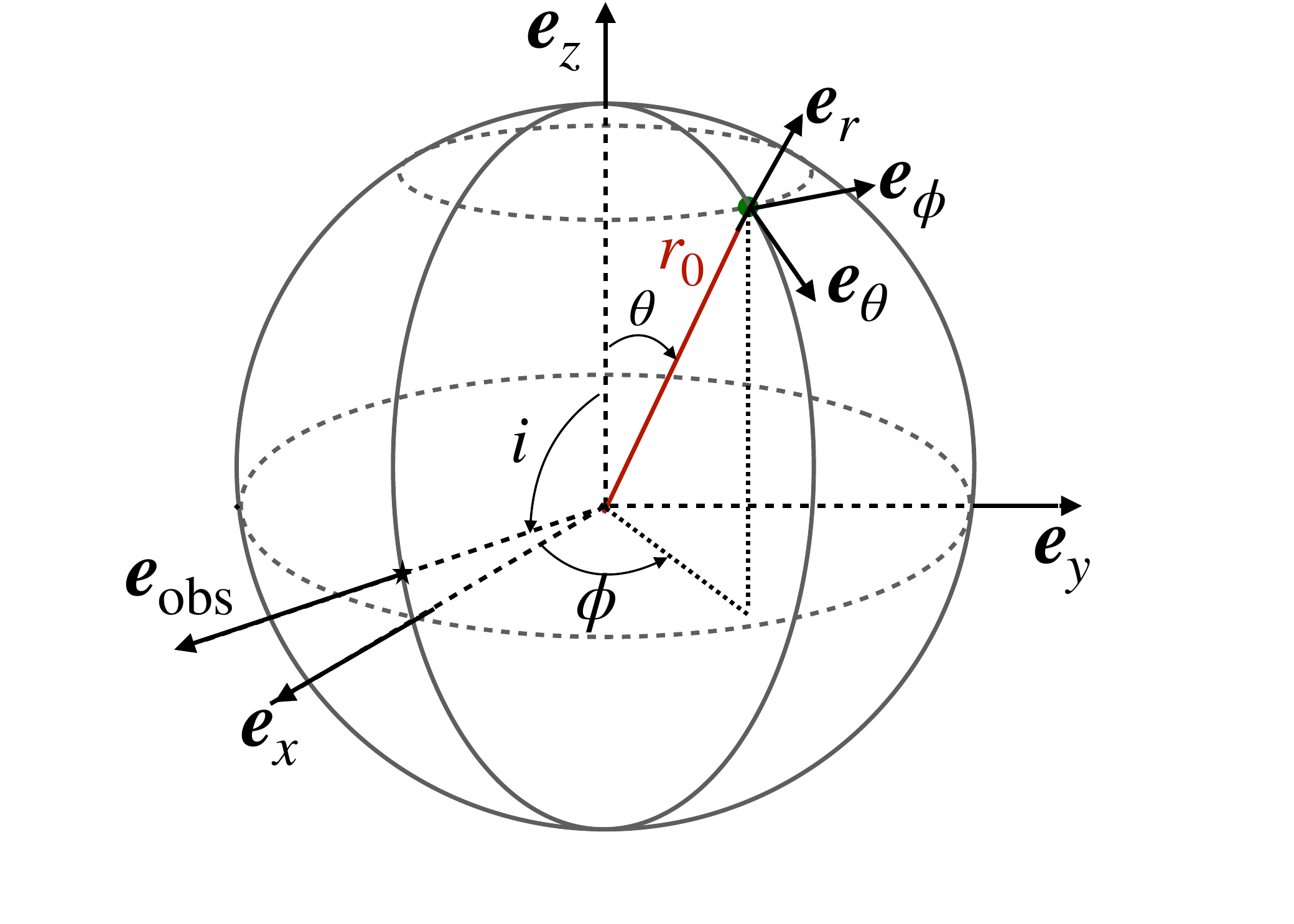}
    \caption{Sketch showing the coordinate systems as well as the different angles used in this study. The reference Cartesian frame is denoted by $(\be_x, \be_y, \be_z)$ and the spherical reference frame by $(\be_r, \be_\theta, \be_\phi)$. The vector $\be_{\rm{obs}}$ is pointing towards the observer. }
    \label{fig:coordinates}    
\end{figure}

As the main purpose of our paper is to establish the link between the wave displacement of different oscillation modes and the emergent intensity perturbations, which are performed in different frames, i.e. inertial and observer, respectively, we first describe these frames and the connection between them. Figure~\ref{fig:coordinates} presents the coordinate systems.
The Cartesian reference (inertial) frame is denoted by $(\be_x,\be_y,\be_z)$ where $\be_z$ is the rotation axis of the Sun. As the Sun rotates slowly, in this paper we neglect its rotation. The spherical unit vectors in the reference frame are denoted by $(\be_r, \be_\theta,\be_\phi)$ with polar and azimuthal angles $\theta$ and $\phi$. The vector $\be_{\textrm{obs}}$ points in the direction of the observer
\begin{equation}
     \be_{\textrm{obs}} = \be_z \cos{i}   +  \be_x \sin{i} ,
     \label{Eq:Eobs_full} 
\end{equation} 
where $i$ is the inclination angle. For the Sun, the inclination angle varying from  $83^\circ$ to $97^\circ$ during the year.
These variations can be responsible for some systematic errors in the data analysis and must be taken into account (\citealt{2018A&A...619A..99L}). It is thus important to keep the inclination angle in the theoretical derivation of intensity perturbations. However, for the numerical tests in this paper, we use $i = 90^\circ$. 

\subsection{Radiative transfer equation} \label{RTE}

The emergent intensity $I(\nu)$ at light frequency $\nu$ is computed at each point with coordinates $(\theta,\phi)$ on the visible hemisphere. To solve the radiative transfer problem we use a plane-parallel approximation, which is valid for most of the positions on the solar disk except very close to the limb.
A comparison of intensity perturbations in plane-parallel and spherical geometry was done by \citet{1999AA...344..188T} for low-degree modes. They showed that the differences between the two geometries become significant only very close to the limb ($\mu=0.1$ corresponding to latitudes higher than $84^\circ$). Therefore, we can assume that our computations are also valid for $0.1 \le \mu \le 1.0$.

\begin{figure}
    \centering
    \includegraphics[width=0.9\linewidth]{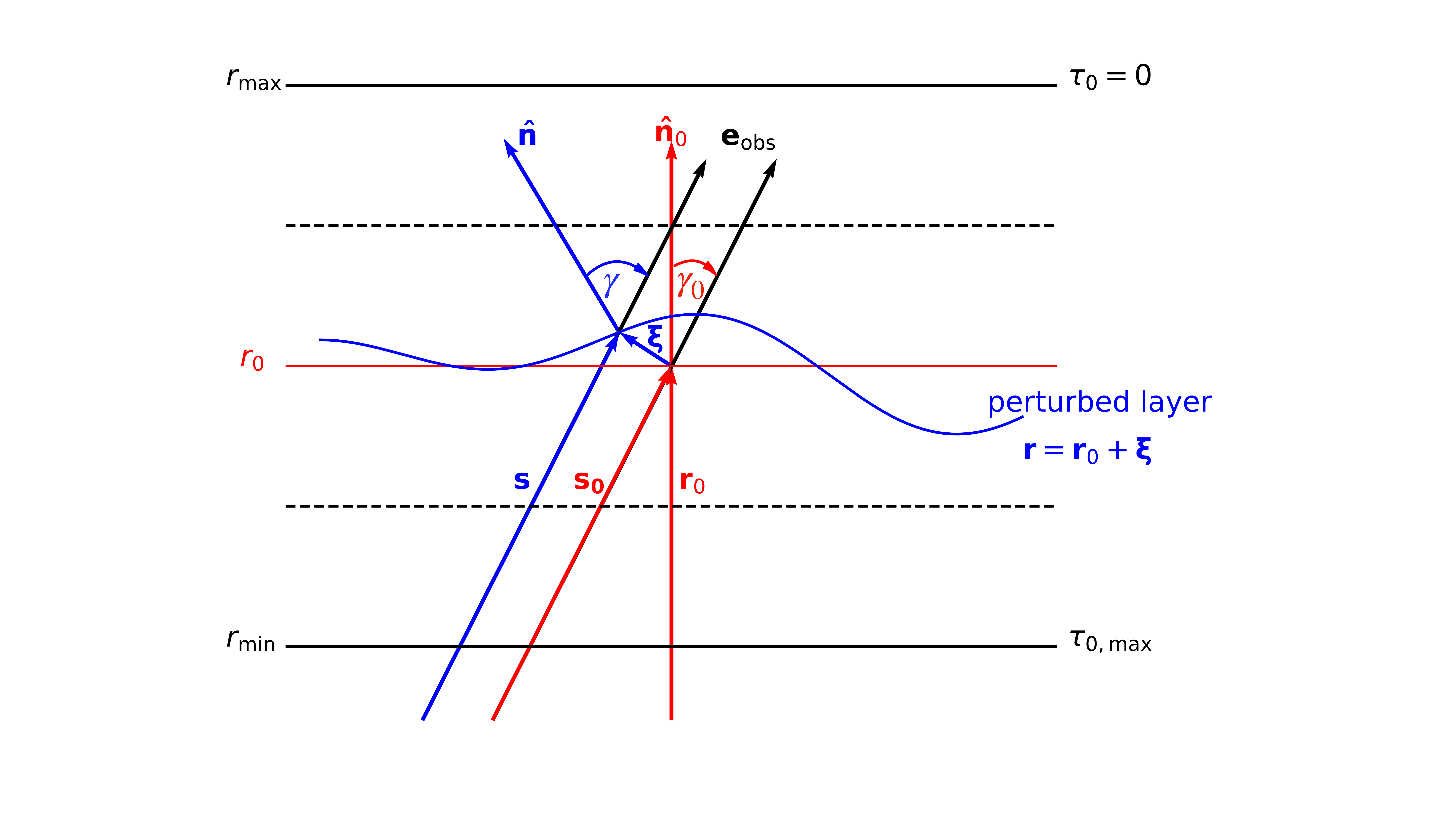}
    \caption{Sketch showing the geometrical quantities involved in solving the radiative transfer equation in the initial and perturbed atmospheres. We zoomed around the atmospheric layers located between $r_{\rm min}$ and $r_{\rm max}$ where 
    the plane-parallel approximation is justified. Red lines and symbols show the quantities in the background model while the perturbed quantities are presented with blue.}
    \label{fig:Sketch}    
\end{figure}

The center-to-limb distance on the disk, $\mu$, is defined as
\begin{equation}
    \mu = \be_{\textrm{obs}} \cdot \hat{\boldsymbol{n}} = \cos\gamma, 
\end{equation}
where $\hat{\boldsymbol{n}}$ is the normal to the solar surface at center-to-limb distance $(r,\theta,\phi)$ and $\gamma$ is the angle between $\hat{\boldsymbol{n}}$ and $\be_{\textrm{obs}}$ as shown on the sketch of Fig.~\ref{fig:Sketch}. 
Here, we assume that the observer is far enough from the Sun, so that the emergent rays at different positions on the disk are always parallel to the direction to the observer. 

The computation of emergent intensity requires integrating the formal solution of the radiative transfer equation (RTE) at frequency $\nu$ along a ray in the direction of the observer over all atmospheric layers
\begin{equation}
	I(\nu, \mu) = \int_{0}^{\infty}S(\nu, s) ~e^{-\tau(\nu, s)/\mu}\frac{\mathrm{d}\tau(\nu, s)}{\mu}.
    \label{intensity}
\end{equation}
Here, the differential of optical depth is defined as

\begin{equation}
	\mathrm{d}\tau(\nu, s) = - \alpha(\nu, s) \mathrm{d} s
    \label{dtau}
\end{equation}

\noindent with $\tau = 0$ at the top of atmosphere (see Fig.~\ref{fig:Sketch}). The extinction coefficient $\alpha(\nu, s)$ describes the total opacity along the ray, and $s$ is the length of the integration path.

The last term to define in Eq.~\eqref{intensity} is the source function $S(\nu, s)$. Assuming local thermodynamic equilibrium which is an adequate approximation in the lower solar photosphere, $S(\nu, s)$ can be expressed as a Planck function
\begin{equation}
	S(\nu, s) = \frac{2\rm{h}\nu^3}{c^2}\frac{1}{e^{{\rm h} \nu/k T}-1},
	\label{eq:sf}
\end{equation}
where $c$ is the speed of light and h and $k$ are the Planck and  Boltzmann constants. 

To avoid solving the complete set of hydrodynamics equations at each point on the solar disk, 
we linearise all perturbed quantities around a background state, which is described by radial coordinate $r_0$, background temperature $T_0$, and pressure $p_0$.
The background intensity $I_0$ is computed at $\mu_0$ defined as cosine of the angle between the observer and the reference normal vector $\hat{\boldsymbol{n}}_0 = \be_r$ (Fig.~\ref{fig:Sketch}),
\begin{equation}
\mu_0 = \be_{\textrm{obs}} \cdot \hat{\boldsymbol{n}}_0 = \cos{\gamma_0} =\sin{i}\sin{\theta}\cos{\phi}+\cos{i}\cos{\theta}.
\label{Eq:mu0}
\end{equation}

\subsection{Perturbations of the path and of thermodynamical quantities of the atmospheric layers due to solar oscillations}

In this subsection we present the perturbed quantities of the model atmosphere. As the surface oscillates the displacement $\br$ fluctuates not only in the radial but in all directions and is written in term of the Lagrangian wave displacement vector $\bxi$
\begin{equation}
    \br = \br_0 + \bxi = (r_0 + \xi_r) \ \be_r + \xi_\theta \ \be_\theta + \xi_\phi \ \be_\phi.
    \label{Eq_r}
\end{equation}
An equivalent decomposition can be written with an Eulerian displacement vector instead of its Lagrangian description. However, \citet{1999AA...344..188T} found that the final expression of the emergent intensity in this framework is computationally challenging since two terms the emission and the absorption, are almost cancelling each other while their difference is the important quantity. Therefore we opted to use a Lagrangian formalism where an expression for the difference is directly obtained.

The wave displacement $\bxi$ changes the length of the integration path across the atmospheric layers $s$ such that
\begin{equation}
    s = s_0 + \delta s,
\end{equation}
where the perturbation $\delta s$ is
\begin{align}
    \frac{\delta s}{s_0} &= \frac{1}{\mu_0 r_0} \  \boldsymbol{\xi} \cdot \be_{\textrm{obs}}, \\
    &= \frac{1}{r_0} \left( \xi_r  + \frac{ \sin{i} \cos{\theta} \cos{\phi} - \cos{i} \sin{\theta}}{\mu_0} \xi_\theta - \frac{\sin{i} \sin{\phi}}{\mu_0} \xi_\phi \right).
    \label{eq:deltaH}
\end{align}
We used the expression of $\be_{\textrm{obs}}$ in the spherical basis given by Eq.~\eqref{eq:eObsSph} to compute the scalar product between $\boldsymbol{\xi}$ and $\be_{\textrm{obs}}$. 

The oscillations also modify the thermodynamical quantities $T$ and $p$. We linearise them around the equilibrium state such that
\begin{equation}
T = T_0 + \delta T; \: p = p_0 + \delta p; \label{eq:firstOrder}
\end{equation}
\noindent where $\delta$ indicates the Lagrangian perturbations of the different quantities. Using the adiabatic approximation, the Lagrangian perturbations of temperature $\delta T$ and pressure $\delta p$ are
\begin{align}
    \frac{\delta T}{T} &=  -\left( \Gamma_3 - 1 \right) \nabla \cdot \bxi, \label{eq:deltaT} \\
    \frac{\delta p}{p} &= - \Gamma_1 \nabla \cdot \bxi,
    \label{eq:deltaP}
\end{align}
where $\Gamma_1$ and $\Gamma_3$ are the first and third adiabatic exponents \citep[Eq.~3.18 in][]{JCD03}. For a neutral and fully ionized hydrogen gas, $\Gamma_3$ approaches $5/3$ and decreases in partially ionised regions, such as below the optical surface where continuum forms or in the lower chromosphere. As in this paper, we focus on the continuum formation region, constant $\Gamma_3$ is not a good approximation, therefore we take a depth-dependent adiabatic exponent from the background model (Sect. \ref{ss:background}). Note, that in this paper, for the results presentation, we suppose adiabaticity for simplicity so that $\bxi$ is the input parameter but the perturbations of all the thermodynamical quantities could also be obtained without this hypothesis by solving the linear nonadiabatic oscillation equations \citep[e.g. Section 13.3 in][]{1989nos..book.....U}.  In order to compare with observations, it is required to take nonadiabaticity into account.

\subsection{Perturbations of the source function, optical depth, and center-to-limb distance}

The perturbations of the length of the integration path and of the thermodynamical quantities modify the source function,  the opacity and thus the optical depth and the center-to-limb distance from their equilibrium values:
\begin{equation}
    S = S_0 + \delta S,  \;  \alpha = \alpha_0 + \delta\alpha, \; \tau = \tau_0 + \delta \tau, \; {\rm and }\ \mu = \mu_0 +\delta \mu.
    \label{eq:pert_smt}
\end{equation}

By perturbing Eq.~\eqref{eq:sf} around the temperature $T_0$, we find that the perturbation to the source function is
\begin{equation}
	\frac{\delta S}{S_0} = \frac{{\rm h}\nu/kT_0}{1-e^{-{\rm h}\nu/kT_0}}\frac{\delta T}{T_0}. 
	\label{eq:deltaS}
\end{equation}

The opacity perturbation is caused by fluctuations in temperature and pressure
\begin{equation}
	\frac{\delta \alpha}{\alpha_0} = \frac{\partial \left( \log \alpha_0 \right)}{\partial \left( \log p_0 \right)}\frac{\delta p}{p_0}+\frac{\partial \left( \log \alpha_0 \right)}{\partial \left(\log T_0 \right)}\frac{\delta T}{T_0} .
	\label{eq:deltaAlpha}
\end{equation}
A similar expression could be written in terms of perturbations in temperature and density, as in \citetalias{1993AA...268..309T} and in \citetalias{1996AA...305L..33Z}, however Eq.~\ref{eq:deltaAlpha} is more convenient for us since the code we use for opacity calculation returns opacity as a function of temperature and pressure. 

The perturbation to the optical depth is 
\begin{equation}
	\frac{d(\delta \tau)}{d\tau_0} = \frac{\delta\alpha}{\alpha_0}+ \frac{d(\delta s)}{ds_0},
	\label{eq:deltaTau}    
\end{equation}
where the opacity perturbation  is given by Eq.~\eqref{eq:deltaAlpha} and $\rm{d}(\delta s)  / \rm{d}s_0$ is obtained by taking the derivative of Eq.~\eqref{eq:deltaH}.

The fluctuation of the center-to-limb distance, $\delta\mu$, is given in Appendix~\ref{Append}:
\begin{align}
    \delta \mu = & \frac{\sin{i}\cos{\theta}\cos{\phi} - \cos{i}\sin{\theta}}{r_0}  \left(\xi_\theta - \frac{\partial \xi_r}{\partial \theta}\right) \nonumber \\ &
    - \frac{\sin{i} \sin{\phi}}{r_0\sin{\theta}}  \left( \xi_\phi \, \sin{\theta} - \frac{\partial \xi_r}{\partial \phi} \right).
    \label{eq:deltaMu}
\end{align}
This expression is consistent  with the one derived by \citet{2013A&A...550A..77R} for rapidly rotating stars.

\subsection{Radiative transfer in perturbed atmosphere} \label{RTE_Perturbed}

Using Eq.~\eqref{eq:pert_smt} in the definition of the emergent intensity (Eq.~\ref{intensity}), we obtain the Lagrangian perturbation of the emergent intensity
\begin{equation}
    \delta I =  I(\mu) - I_0(\mu_0)  = \delta I_{S} + \delta I_{\tau} + \delta I_{\mu}, 
    \label{eq:pert_int}
\end{equation}
\noindent where 
\begin{align}
    \delta I_{S} &= \int_{0}^{\tau_{0,{\rm max}}} S_0 e^{-{\tau_0}/{\mu_0}} \frac{\delta S}{S_0}\frac{\mathrm{d}\tau_0}{\mu_0},
    \label{eq:deltaIS} \\
    \delta I_{\tau} &= \int_{0}^{\tau_{0,{\rm max}}}S_0e^{-{\tau_0}/{\mu_0}}\left(\frac{\mathrm{d}\delta\tau}{\mathrm{d}\tau_0} -\frac{\delta \tau}{\mu_0}\right)\frac{\mathrm{d}\tau_0}{\mu_0}, 
    \label{eq:deltaITau} \\
    \delta I_{\mu} &= \int_{0}^{\tau_{0,{\rm max}}}S_0e^{-{\tau_0}/{\mu_0}}\frac{\delta \mu}{\mu_0}\left(\frac{ \tau_0}{\mu_0}-1 \right)\frac{\mathrm{d}\tau_0}{\mu_0}.
    \label{eq:deltaIMu}
\end{align}
The perturbations of the source function $\delta S$, optical depth $\delta \tau$, and incident angle $\delta \mu$ are given respectively by Eqs.~\eqref{eq:deltaS},~\eqref{eq:deltaTau},~and~\eqref{eq:deltaMu}. Note, that in this decomposition of the intensity perturbation, the terms $\delta I_\tau$ and $\delta I_\mu$ both contain contributions from $\boldsymbol{\xi}$.  

\subsection{Comparison of intensity perturbation derivation with previous studies}

In order to compare the intensity perturbation to other studies, i.e. \citetalias{1993AA...268..309T} and \citetalias{1996AA...305L..33Z}, it is more convenient to decompose $\delta I$ into a thermodynamical term ($\delta I_{\rm th}$) which contain all the components with temperature and pressure perturbations and a geometrical term with wave displacement contribution ($\delta I_{\boldsymbol{\xi}}$):
\begin{equation}
    \delta I =  \delta I_{\rm th} + \delta I_{\boldsymbol{\xi}}, 
    \label{eq:pert_int2}
\end{equation}
\noindent where 
\begin{equation}
  \delta I_{\rm th} = \delta I_S + \delta I_{\tau, \,\alpha}.
\end{equation}
 The term $\delta I_{S}$ (Eq.~\ref{eq:deltaIS}) is the same for all three studies, i.e this paper, \citetalias{1993AA...268..309T} and \citetalias{1996AA...305L..33Z}. The contribution of opacity to the emergent intensity perturbation $\delta I_{\tau, \,\alpha}$ is
\begin{equation}
    \delta I_{\tau, \,\alpha} = \int_{0}^{\tau_{0,{\rm max}}}S_0e^{-{\tau_0}/{\mu_0}}\left(\frac{\delta\alpha}{\alpha_0} - \int_0^{\tau_0} \frac{\delta\alpha}{\alpha_0}\frac{\mathrm{d}\tau_0'}{\mu_0} \right)\frac{\mathrm{d}\tau_0}{\mu_0},
    \label{eq:deltaIAlpha}
\end{equation}
    with $\delta \alpha/\alpha_0$ defined in Eq~\eqref{eq:deltaAlpha}.
    
\noindent The contribution from the geometrical term is
    \begin{align}
    \delta I_{\boldsymbol{\xi}} = & \int_{0}^{\tau_{0,{\rm max}}}S_0e^{-{\tau_0}/{\mu_0}} \frac{r_0}{\mu_0} \Biggl\{  
    \frac{\partial }{\partial r_0} \left( \frac{\boldsymbol{\xi} \cdot \be_{\textrm{obs}}}{r_0} \right)
    \nonumber \\ & 
    +  \left( \frac{\tau_0}{\mu_0}-1 \right) \left( \mu_0 \frac{\partial}{\partial r_0} \left( \frac{\xi_r}{r_0} \right) - \frac{\nabla \boldsymbol{\xi} \cdot \be_{\mathrm{obs}}}{r_0}  \right) \Biggr\} \frac{\mathrm{d}\tau_0}{\mu_0}.
    \label{eq:deltaIXi}
\end{align}

As it was mentioned in the introduction, \citetalias{1996AA...305L..33Z} neglected the geometrical effect and took only perturbations of thermodynamical quantities as a source of intensity fluctuations. Applying an integration by part to Eq.~\ref{eq:deltaIAlpha}, the $\delta I_{\rm{th}}$ is then identical to Eq.~1 in \citetalias{1996AA...305L..33Z}.

The last term in Eq.~\eqref{eq:pert_int2}, $\delta I_{\boldsymbol{\xi}}$, contains all the contributions due to $\boldsymbol{\xi}$ and describes purely geometrical effects which are induced by the deformations of the atmospheric layers as well as the center-to-limb distances due to acoustic oscillations. Keeping only the radial displacement $\xi_r$ in Eq.~\ref{eq:deltaIXi}, while neglecting $\xi_\theta$ ($\xi_\theta  \ll \xi_r$) leads to the definition of emergent intensity $\delta I_{\xi_r}$ as in \citetalias{1993AA...268..309T}.
Note that additionally \citetalias{1993AA...268..309T} assumed that  $\xi_r/r$ is not varying with height, while we take the height dependence of $\xi_r/r$ into account. Therefore, our expression of $\delta I_{\textrm{th}} + \delta I_{\xi_r}$ slightly differs from Eq.~4 in \citetalias{1993AA...268..309T}, however, it agrees with the plane-parallel expression $\Delta I^{//}$ derived later by \citet{1999AA...344..188T} where the height-dependence of $\xi_r/r$ has been taken into account.

\section{Numerical inputs for intensity calculations} \label{sect:num}

The acoustic oscillations of the Sun modify its stratification and thus the emergent intensity. The reference intensity is computed in a background model given in Sect.~\ref{ss:background}. We calculate the perturbations caused by a single acoustic mode whose computation is explained in Sect.~\ref{Displacement}. 
In order to determine equilibrium and perturbed intensities, we explain the computation of opacity and its derivatives in Sect.~\ref{sect:opacity}. 

\subsection{Background model}\label{ss:background}

For the background quantities, we use the model S \citep{1996Sci...272.1286C} which uses the OPAL equation of state \citep{1996ApJ...456..902R} and OPAL opacities in the deep layers \citep{1992ApJ...397..717I} and \citet{1991ASIC..341..441K} opacities in the atmosphere. This model is accurate in  the solar interior, however, it is too simplified in the superadiabatic layer close to the surface. The model of convection is based on the mixing-length theory \citep{1958ZA.....46..108B} therefore simplifying the computation of the turbulent pressure which contributes significantly to the emitted radiation in this layer. Due to this simplification, the eigenfrequencies of the solar spectrum do not match the observed ones (surface effect, \citealt{1999A&A...351..689R}). A better agreement with observations is obtained by replacing the background in the atmosphere by averaged quantities coming from numerical simulations \citep{2016A&A...592A.159B} or by patching the eigenfunctions directly calculated from a 3D hydrodynamical simulations onto the one from a 1D model \citep{2020A&A...638A..51S}. 
In order to avoid the difficulties due to the matching of all background quantities between model S and the atmospheric model, we use only model S up to 500~km above the solar surface. This height is sufficient to model continuum intensity. However, the influence of the background model on the intensity should be studied before interpreting the observations.

\subsection{p-mode eigenfunctions} \label{Displacement}

\begin{figure}[ht]
    \centering
    \includegraphics[width=0.99\linewidth]{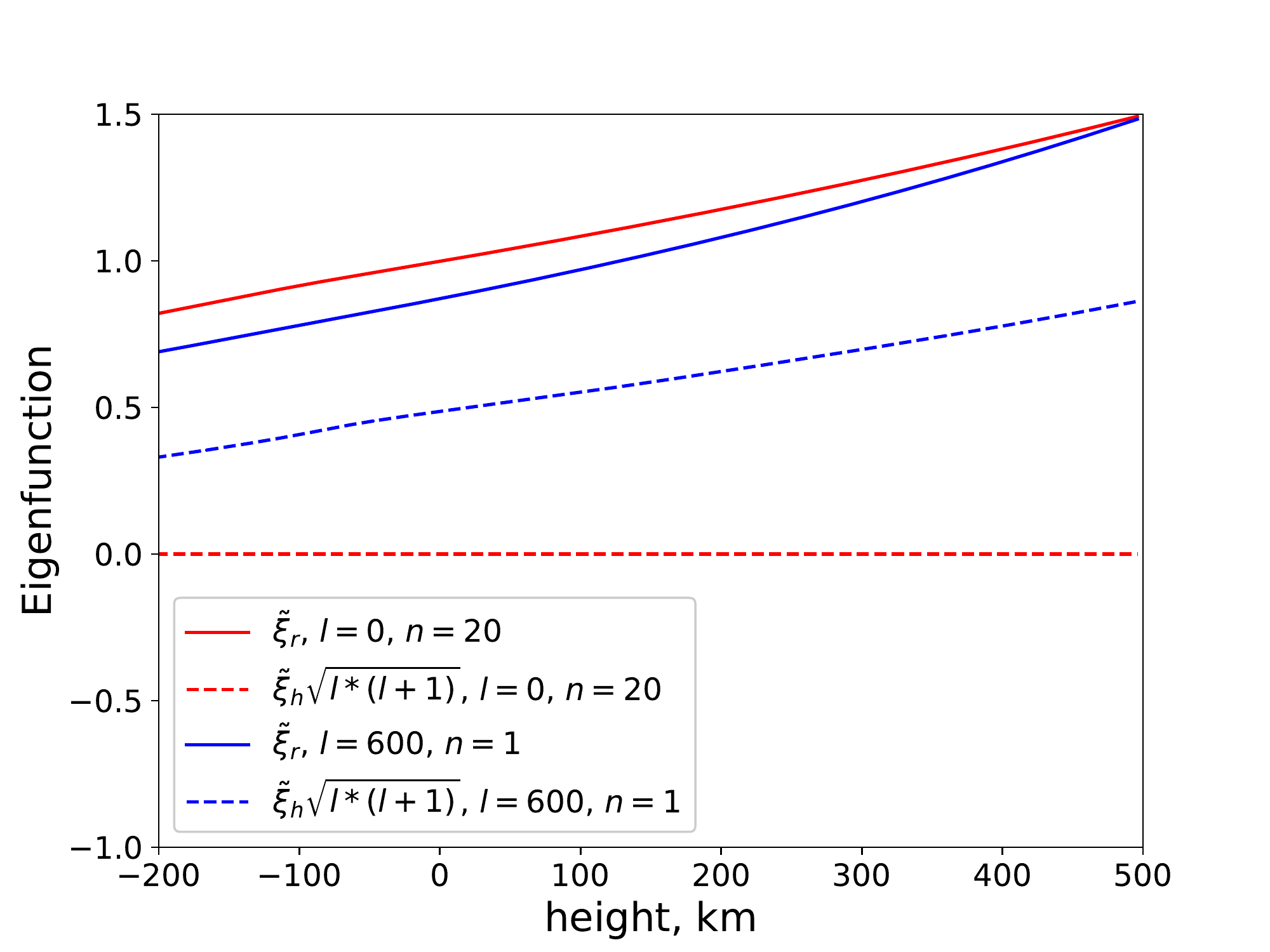}
    \caption{Radial and horizontal eigenfunctions calculated for two normal modes, radial ($l=0$, $m=0$, $n=20$) and non-radial ($l=600$, $m=0$, $n=1$), normalised to the the norm of the displacement $\sqrt{\Tilde{\xi}_r^2 +l(l+1) \ \Tilde{\xi}_h^2}$ at the surface.}
    \label{fig:EigenFunction}
\end{figure}

The normal modes of acoustic oscillations are computed using the ADIPLS code \citep{JCD08}. The displacement vector of non-radial modes in the reference frame is written as \citep[see e.g.][]{JCD03}
\begin{equation}
    \boldsymbol{\xi}(r,\theta,\phi,t) = \xi_r \ \be_r + \xi_\theta \ \be_\theta + \xi_\phi \ \be_\phi,
\end{equation}
where 
\begin{align}
    \xi_r &= \textrm{Re} \left\{ \Tilde{\xi}_r(r) Y_l^m(\theta, \phi) \exp(-\ii \omega t) \right\}, \\
    \xi_\theta &= \textrm{Re} \left\{ \Tilde{\xi}_h(r)  \frac{\partial Y_l^m(\theta, \phi)}{\partial \theta}  \exp(-\ii \omega t) \right\}, \\
    \xi_\phi &= \textrm{Re} \left\{  \Tilde{\xi}_h(r)  \frac{1}{\sin \theta}\frac{\partial Y_l^m(\theta, \phi)}{\partial \phi} \exp(-\ii \omega t) \right\}.
\end{align}
Here, $Y_l^m$ are the spherical harmonics of degree $l$ and azimuthal order $m$. The code solves an eigenvalue problem in a 1D standard solar model \citep[in this paper, model~S,][]{JCD03} to determine the radial $\Tilde{\xi}_r$ and horizontal $\Tilde{\xi}_h$ eigenfunctions associated to the eigenvalue $\omega_{nlm}$. The surface boundary condition is applied 500~km above the solar surface by supposing an isothermal atmosphere \citep{JCD08}. 
For adiabatic oscillations without attenuation, the frequency and the eigenfunctions $\Tilde{\xi}_r$, $\Tilde{\xi}_h$ are real. The variations of a radial and a  non-radial mode with height are shown on Fig.~\ref{fig:EigenFunction}. For high-degree modes with eigenfrequency around $3~\rm mHz$, the horizontal part $\Tilde{\xi}_h$ becomes comparable in amplitude to the radial part $\Tilde{\xi}_r$ and justifies that we have kept the horizontal displacements $\xi_\theta$ and $\xi_\phi$ in our derivation of intensity perturbations. Table \ref{tab:modes} lists the exact values of eigenfrequencies ($\omega_{lmn}/2\pi$) considered in this paper which have been chosen around $3~\rm mHz$ corresponding to the 5-min solar oscillations. For each mode, we give the ratio between the horizontal and radial displacements ($\sqrt{l(l+1)} \Tilde{\xi}_h / \Tilde{\xi}_r$) at the surface showing the importance of the horizontal displacement for each of the modes ($l$, $m$, $n$) with frequency $\omega_{lmn}/2\pi$.

\begin{table}
    \centering
    \caption{Acoustic modes with their eigenfrequencies and ratio of horizontal-to-radial components at the surface (at solar radius of displacement computed using ADIPLS \citep{JCD08}.}
    \begin{tabular}{c|c|c|c|c}
        $l$ & $m$ & $n$ & $\omega_{lmn}/2\pi$ [mHz] & $\sqrt{l(l+1)} \Tilde{\xi}_h  / \Tilde{\xi}_r$ \\
        \hline
        0 & 0 & 20 & 2.902 & 0 \\ 
        100 & 0 & 6 & 2.936 & 0.094 \\ 
        600 & 0 & 1 & 3.043 & 0.559 \\ 
    \end{tabular}
    \label{tab:modes}
\end{table}

\subsection{Opacity} \label{sect:opacity}

To compute the optical depth along which the RTE is solved, the opacity as a function of depth should be known. We compute absorption and scattering coefficients using the Merged Parallelized Simplified ATLAS code (MPS-ATLAS, \citealt{Witzke}) developed from the original ATLAS code \citep{Kurucz_manual_1970}. To compute continuum opacity we take into account the contributions from the bound-free and free-free transitions of $\rm H^-$, $\rm HI$, $\rm H_2^+$, the free-free transitions of $\rm He^-$, metal photo ionization, Rayleigh scattering on $\rm HI$ and $\rm He I$, and Thomson scattering on free electrons. Those are the main sources of opacity in the continuum.

In addition to the opacity in the background model, we also need the derivatives $\left(\partial  \ln \alpha_0 /{\partial \ln p }\right)|_{T_0}$ and $\left({\partial\ln \alpha_0}/{\partial\ln T}\right)|_{p_0}$ in order to compute the perturbation of opacity (Eq.~\eqref{eq:deltaAlpha}). 
Taking into account only the bound-free transition of $\rm H^-$, which is the main source of continuum opacity in the visible wavelength range, \citetalias{1993AA...268..309T} presented the analytic equation of the opacity derivative. However, as it shown in \citet{2015AA...575A..89K}, the bound-free transition of $\rm H^-$ is not the only contributor to the total opacity and the contribution from other sources of opacity can be larger than $\rm H^-$ at some heights in the solar atmosphere. Adding other contributors to the continuum opacity makes the derivation of analytical expression intractable.
Another approach is to compute the derivatives of opacity taking all possible contributors into account using the pre-computed opacity table on some temperature-pressure (or density) grid.
However, this requires interpolation in order to evaluate the derivatives at the temperature and pressure (density) of the model atmosphere. This approach was first applied by \citet{1995ASPC...76..338S} in the VALC3 model of atmosphere from \citep{1981ApJS...45..635V}. Later, \citetalias{1996AA...305L..33Z} showed that this approach leads to non-smooth derivatives and additionally applied the fitting and smoothing procedure as in OPFIT \citep{1994MNRAS.266..805S}. They showed that the intensity fluctuations caused by the same oscillation mode and at the same wavelength are slightly different from \citet{1995ASPC...76..338S}, who did not use any smoothing. In order to avoid any interpolation, fitting, and smoothing schemes that may introduce additional uncertainties to the intensity perturbations, we compute our opacity table for the Model S grid of temperature and pressure using the MPS-ATLAS code. From the table, we calculate numerically the partial derivatives of the opacity. 

\section{Computation of intensity perturbation due to acoustic oscillations} \label{results}

In this section, in order to validate our algorithm, we compare the perturbed intensity computed using the algorithm described in Section~\ref{s:IntPert} with the intensity computed in a perturbed atmosphere directly. Then we present the computation of intensity perturbation caused by p-mode oscillations of different harmonic degrees in the stratified model atmosphere. As we do not study spectral dependence of intensity perturbation in this paper all calculations are done at $500~\rm nm$. The radial orders of the considered oscillation modes are chosen such that the frequency of the oscillations is around $3~\rm mHz$ (Table~\ref{tab:modes}).

\subsection{Comparison with direct computation}

As was mentioned in Section~\ref{s:IntPert}, we derive an emergent intensity in the oscillating atmosphere assuming first-order approximation for the perturbations caused by these oscillations. To validate this approximation, we compare the intensity in a perturbed model $ I(\mu)$ to the sum of the intensity in an initial model plus a perturbation $I_0(\mu_0) + \delta I$. For the test calculation, we assume that the perturbation is caused by the radial mode ($l = 0$, $m = 0$, $n = 20$). 

\begin{figure}[!htb]
    \centering
    \includegraphics[width=0.99\linewidth]{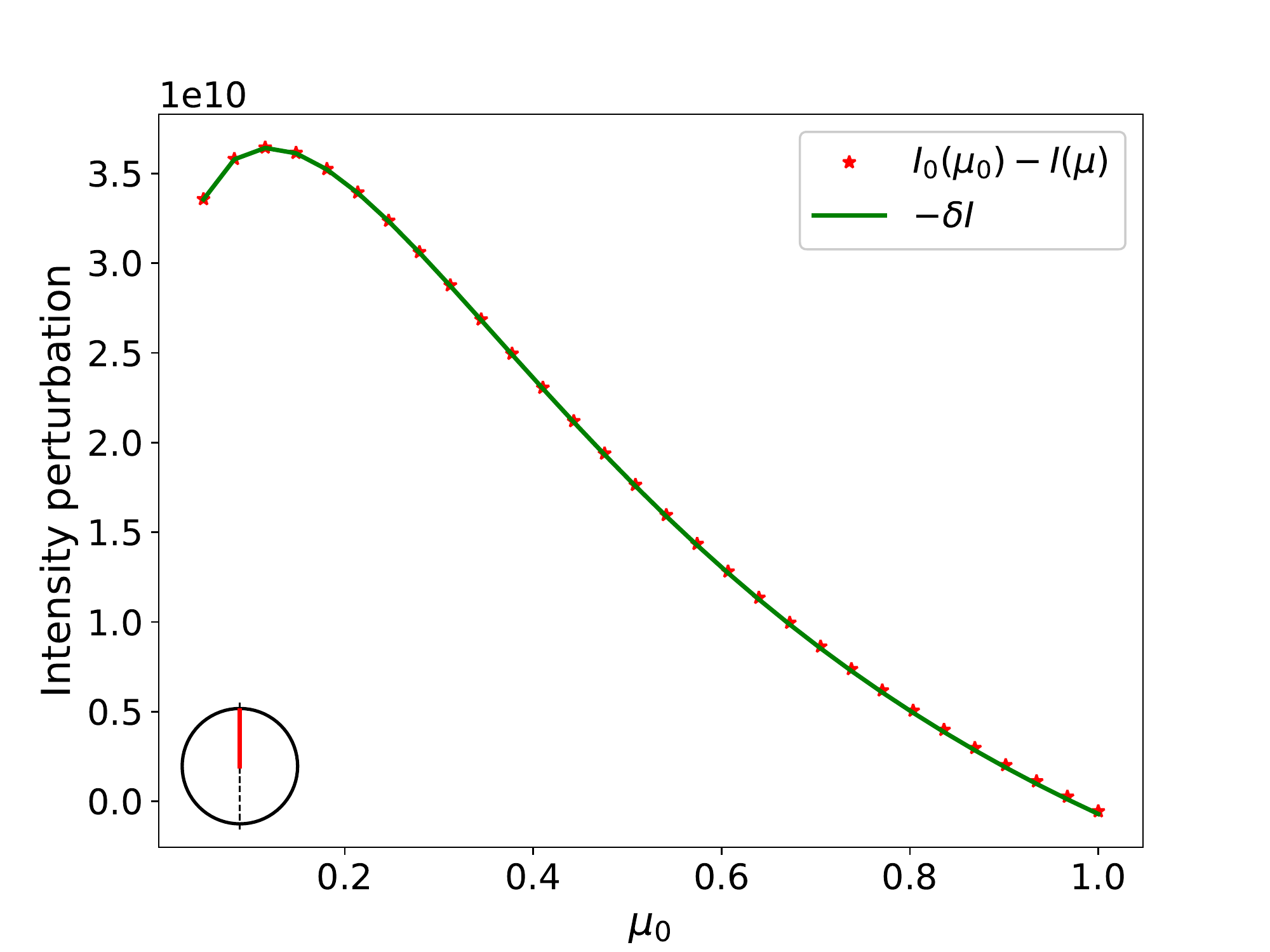}
    \caption{Comparison of $\delta I$ computed from the first-order perturbation theory to the difference of direct computation of emergent intensity in the initial and perturbed model atmospheres $I_0(\mu_0) - I(\mu)$. Intensities are calculated from the center to the limb for the radial mode $l = 0$, $m = 0$, $n = 20$. The red line on the panel in the left corner shows the positions on the solar disk in the observer frame where intensity perturbations are computed and which corresponds to $i=90^\circ$, $\phi=0$ and $ 0 \le \theta < 90^\circ$. The intensity perturbation is measured in $[\rm{erg}~\rm{cm^{-2}}\rm{s^{-1}}\rm{cm^{-1}}\rm{sr^{-1}}]$.}
    \label{fig:test}
\end{figure}

We compute $I_0(\mu_0)$ with $T_0$ and $p_0$ at $s_0$ coming from model~S. We then perturb the model by adding $\delta T$, $\delta p$ and $\delta s$ coming from the eigenfunction of a radial mode multiplied by a factor of $10^4$ and follow the same procedure to compute  $I(\mu)$ in a model characterized by $T = T_0 + \delta T$, $p = p_0 + \delta p$ and $s = s_0 + \delta s$. The last step is to compute the intensity perturbation $\delta I$ due to $\delta T$, $\delta p$ and $\delta s$ by applying the first-order perturbation theory described in Sect.~\ref{RTE_Perturbed}. Note that the factor we multiplied of the mode is only important for the direct computation as we need to detect the response of intensity on the caused perturbations. The factor should be large enough to be visible in the direct computation of intensity but not too large so that the first-order becomes invalid.
Figure~\ref{fig:test} shows that the perturbed intensity $\delta I$ coincides with the difference of direct computations of intensity $I(\mu)$ in the perturbed atmosphere and the intensity $I_0(\mu_0)$ in the reference background model. Thus, this agreement between both approaches allows us to conclude that the first-order perturbation theory is applicable (even for the large perturbation used in this test) and the considered algorithm is correctly implemented. 

\subsection{Comparison of intensity computation in other studies} \label{sect:low}

The intensity perturbations caused by radial p-modes were studied by \citetalias{1993AA...268..309T} and \citetalias{1996AA...305L..33Z} but showed slightly different results. On one hand, \citetalias{1996AA...305L..33Z} claimed that the inconsistency happened because \citetalias{1993AA...268..309T} considered only absorption by $\rm H^-$ as source of opacity. On the other hand, \citetalias{1996AA...305L..33Z} considered only perturbations of the thermodynamical quantities and neglected the geometrical effect. 
Here, we investigate these differences and we thus compute intensity perturbations ($\delta I_{\rm th} + \delta I_{\xi_r}$) taking the main sources of opacities in the continuum into account (Sect.~\ref{sect:opacity}),  $\delta I_{\rm th}$ as in \citetalias{1996AA...305L..33Z}, and $\delta I_{\rm {H^-}}$ with only $\rm H^-$ contribution to the opacity as in \citetalias{1993AA...268..309T}.
In all cases, we compute opacity tables and their derivatives numerically and do not use the analytic expressions from \citetalias{1993AA...268..309T}.  We select the same radial mode as in \citetalias{1993AA...268..309T} and \citetalias{1996AA...305L..33Z} and solve RTE at the same wavelength, however our models of atmosphere are different so we cannot compare directly our results with these papers. Nevertheless, our analysis will help us to understand the differences between the different simplifications and which terms are important when evaluating intensity perturbations.  

\begin{figure}[!htb]
    \centering
    \includegraphics[width=0.99\linewidth]{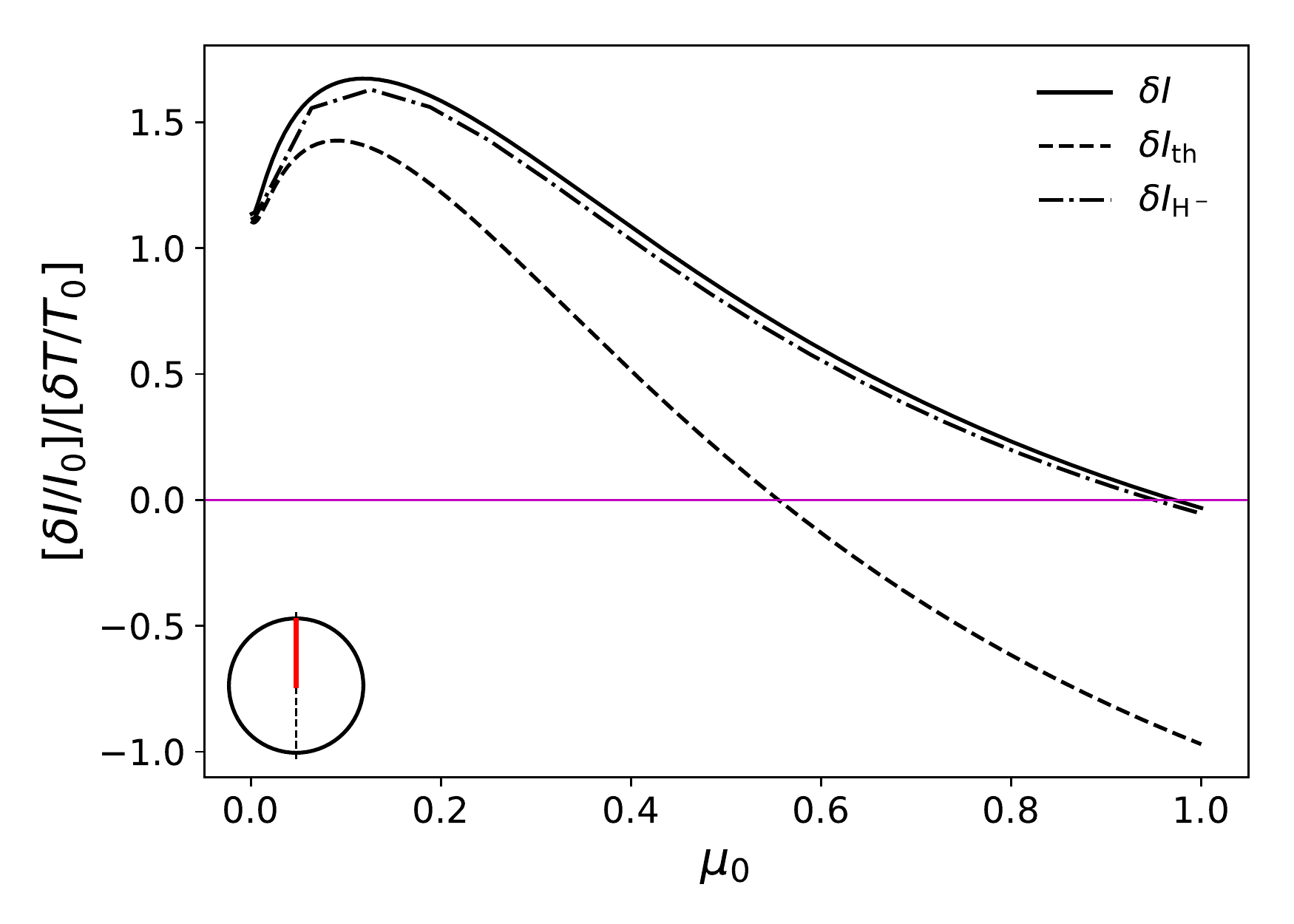}
    \caption{Normalized emergent intensity perturbation caused by the radial mode ($l=0$, $m=0$, $n=20$). The normalization factor $\delta T/T_0$ is taken at $\tau_0 = 1$. The intensity perturbations are computed at the same points on the solar disk as in Fig.~\ref{fig:test}, i.e., $i=90^\circ$, $\phi=0$ and $ 0 \le \theta < 90^\circ$ (see observer view in the lower left corner).}
    \label{fig:RadialMode}
\end{figure}

Fig.~\ref{fig:RadialMode} shows the perturbed emergent intensity divided by the reference intensity $(\delta I / I_0)$ at different $\mu$ normalized by the temperature variations $(\delta T / T_0)$ at the optical surface where $\tau = 1$. The first-order approach is linear, so the amplitude of the mode is cancelled out by such a normalization. As it is not trivial to get the amplitude of each mode, such a normalization allows us to avoid this difficulty. So, all figures below present the normalized emergent intensity.

Moreover, this normalization allows us to understand whether approximating the continuum intensity observable by the temperature perturbation at the optical surface is good enough. If $\delta T / T_0$ is a good approximation of $\delta I / I_0$ then we should see all three normalised intensity curves as a horizontal line at i.e. $[\delta I / I_0]/[\delta T / T_0] = 4$ for the black-body approximation with no variation on $\mu$, which is clearly not the case (Fig.~\ref{fig:RadialMode}). The largest difference is between $\delta I$ and $\delta I_{\rm th}$ which comes from neglecting the geometrical terms in the latter. This is in contradiction with the intuition of \citetalias{1996AA...305L..33Z} who assumed that the surface distortion was not significantly affecting the emergent intensity for low-degree modes. To further analyse the contribution of different opacity sources to the emergent intensity calculation, we compare the intensity $\delta I_{\rm H^{-}}$ with $\delta I$ as both of them contain the geometrical terms. The two curves are very similar with minor deviations along the disk showing that bound-free and free-free transitions of $\rm H^-$ are the main but not the only sources of opacity that contribute to the emergent intensity computation.  
Therefore, neglecting other sources of continuum opacity at 500~nm have little influence of the continuum intensity and could not explain the divergence between \citetalias{1993AA...268..309T} and \citetalias{1996AA...305L..33Z} which arises due to the geometrical effects. 

\subsection{Comparison of the different contributions to the perturbation of emergent intensity.}

\begin{figure*}[!htb]
    \centering
    \includegraphics[width=0.39\linewidth]{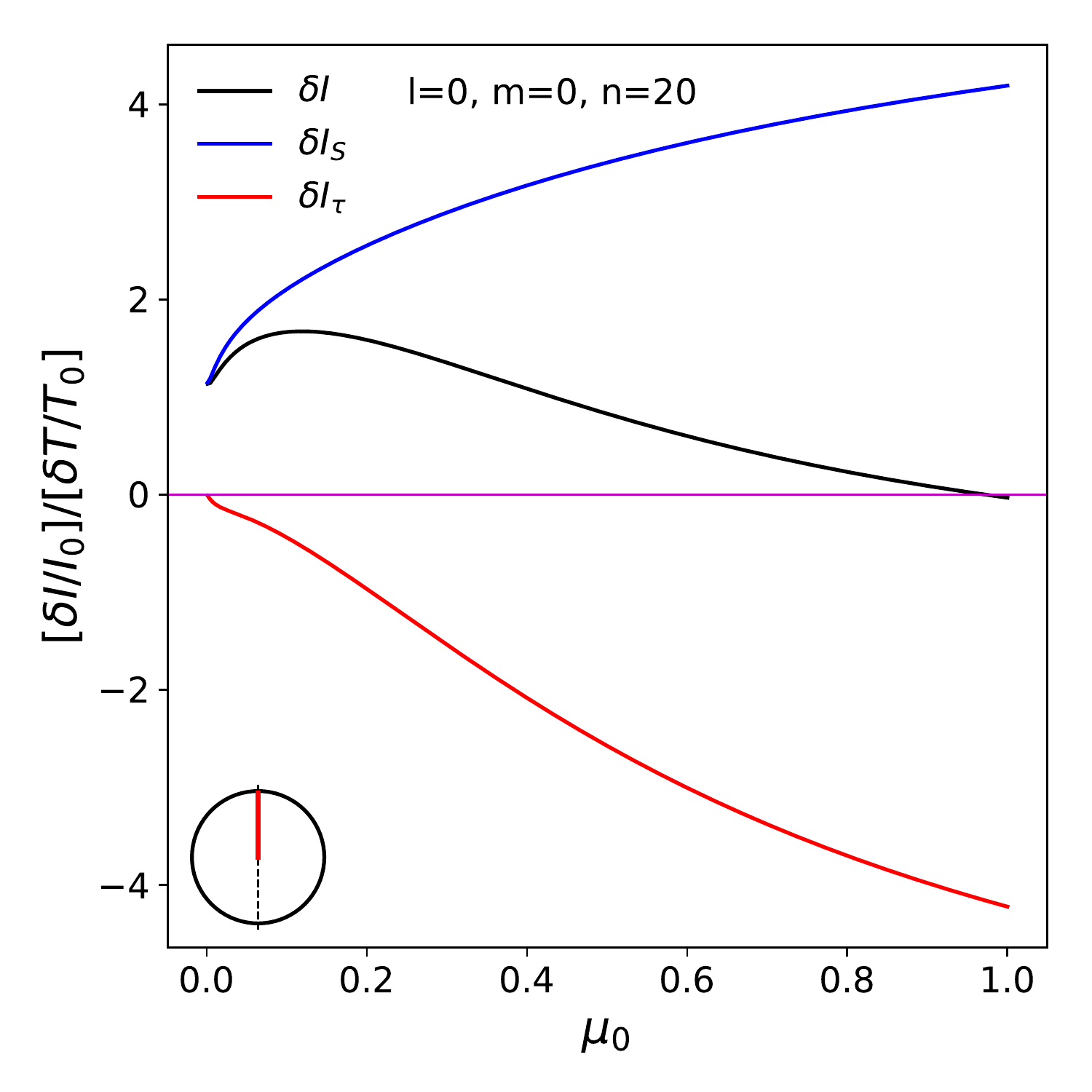}
    \includegraphics[width=0.59\linewidth]{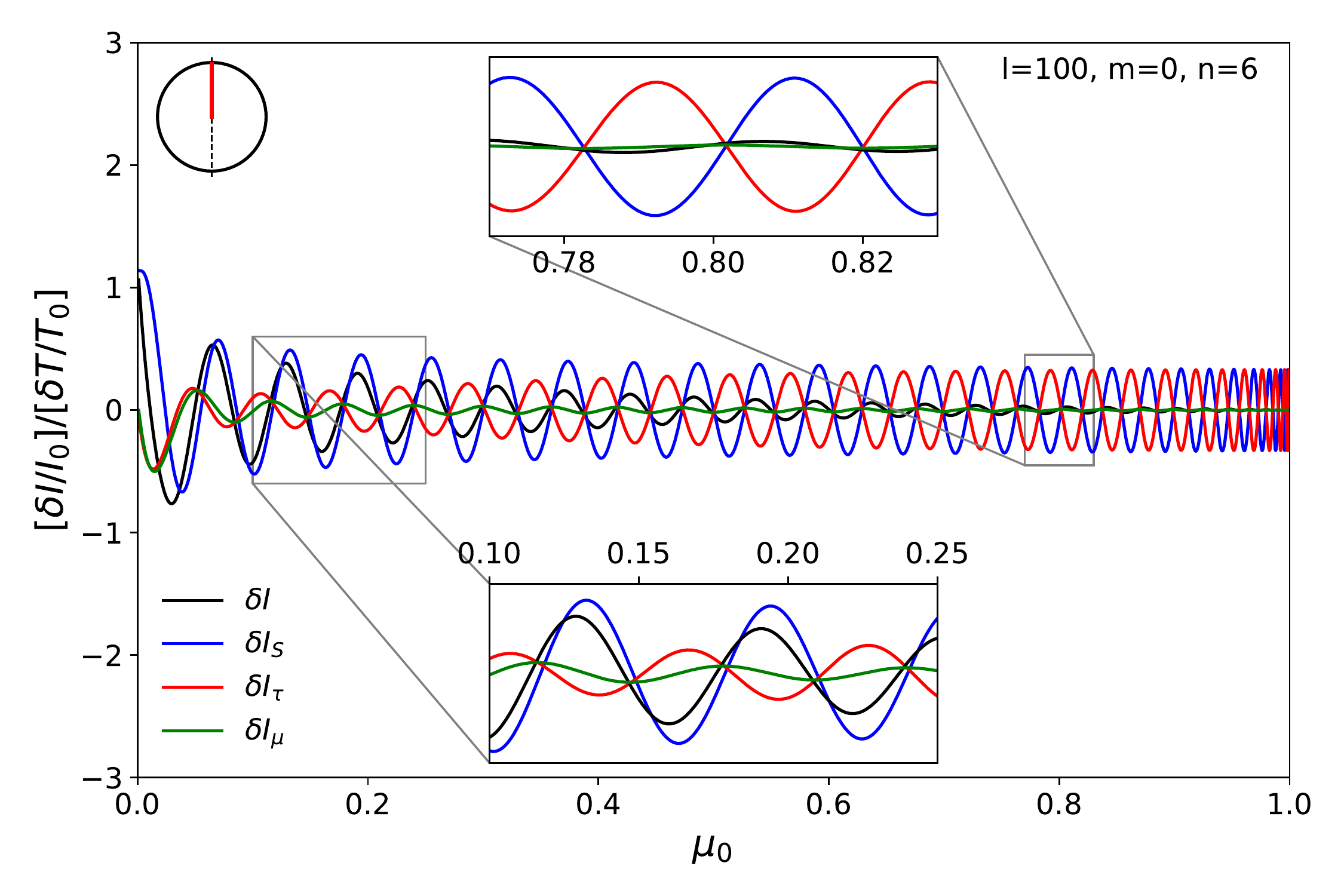}
    \caption{Normalised intensity perturbation and its contributions caused by the radial ($l=0$, $m=0$, $n=20$) and moderate degree ($l=100$, $m=0$, $n=6$) modes. The normalization factor $\delta T/T_0$ is taken at $\tau_0 = 1$. The intensity perturbations are computed at the same points on the solar disk as in Fig.~\ref{fig:test}, i.e., $i=90^\circ$, $\phi=0$ and $ 0 \le \theta < 90^\circ$ (see observer view on the upper, or lower left corner).}
    \label{fig:RadialMode_Contibutors}
\end{figure*}

Perturbations of emergent intensity caused by oscillation modes are coming from three contributors describing the radiative transfer in the solar atmosphere (Eqs.~ \ref{eq:deltaIS},~\ref{eq:deltaITau}, and \ref{eq:deltaIMu}). In this subsection, we study how each of these components affects the emergent intensity perturbation caused by a radial mode as well as a intermediate degree mode ($l=100$) which is observed with spatially resolved instruments (Table~\ref{tab:modes}). In Fig.~\ref{fig:RadialMode_Contibutors}, we show that the emergent intensity $\delta I$ is a balance between $\delta I_S$ and $\delta I_\tau$ for the $l=0$ mode, while for the l=100 mode in addition to the $\delta I_S$ and the $\delta I_\tau$ components, the contribution from $\delta I_\mu$ becomes significant especially close to the limb. The amplitudes of  $\delta I_S$ and $\delta I_\tau$ are similar at the disk center ($\mu=1$) with opposite sign and decrease towards the limb ($\mu=0$) where most of the radiation comes from the higher layers with lower temperature and pressure. Both components thus decrease and as the optical depth drops exponentially, the value of $\delta I_\tau$ decreases faster than $\delta I_S$. The negative sign of $\delta I_\tau$ is because $\frac{d (\delta \tau)}{d \tau} - \frac{\delta \tau}{\mu_0} < 0$  in Eq.~\ref{eq:deltaITau}. For the $l=100$ mode the horizontal displacement contributes to a phase shift of $\delta I_\tau$ (Eq.~\ref{eq:deltaH}) with respect to $\delta I_S$ especially close to the limb.
Thus the intensity perturbation due to the opacity and source function perturbations almost compensate each other for the radial mode and close to the disc center for the moderate degree modes but these two components are phase shifted for observation close to the limb. 

Additional phase shift comes from $\delta I_\mu$, which becomes more and more important as the degree of the mode increases and is zero for the radial modes.
As the emergent intensity combine all the terms, $\delta I$ is shifted with respect to $\delta I_S$, and thus $\delta T$. Therefore, this effect can lead to some systematic errors in intensity maps analysis in helioseismology when the radiative transfer is neglected.

\subsection{Importance of the geometrical terms with horizontal displacement contribution for intensity computation}

\begin{figure*}[!htb]
    \includegraphics[width=0.49\linewidth]{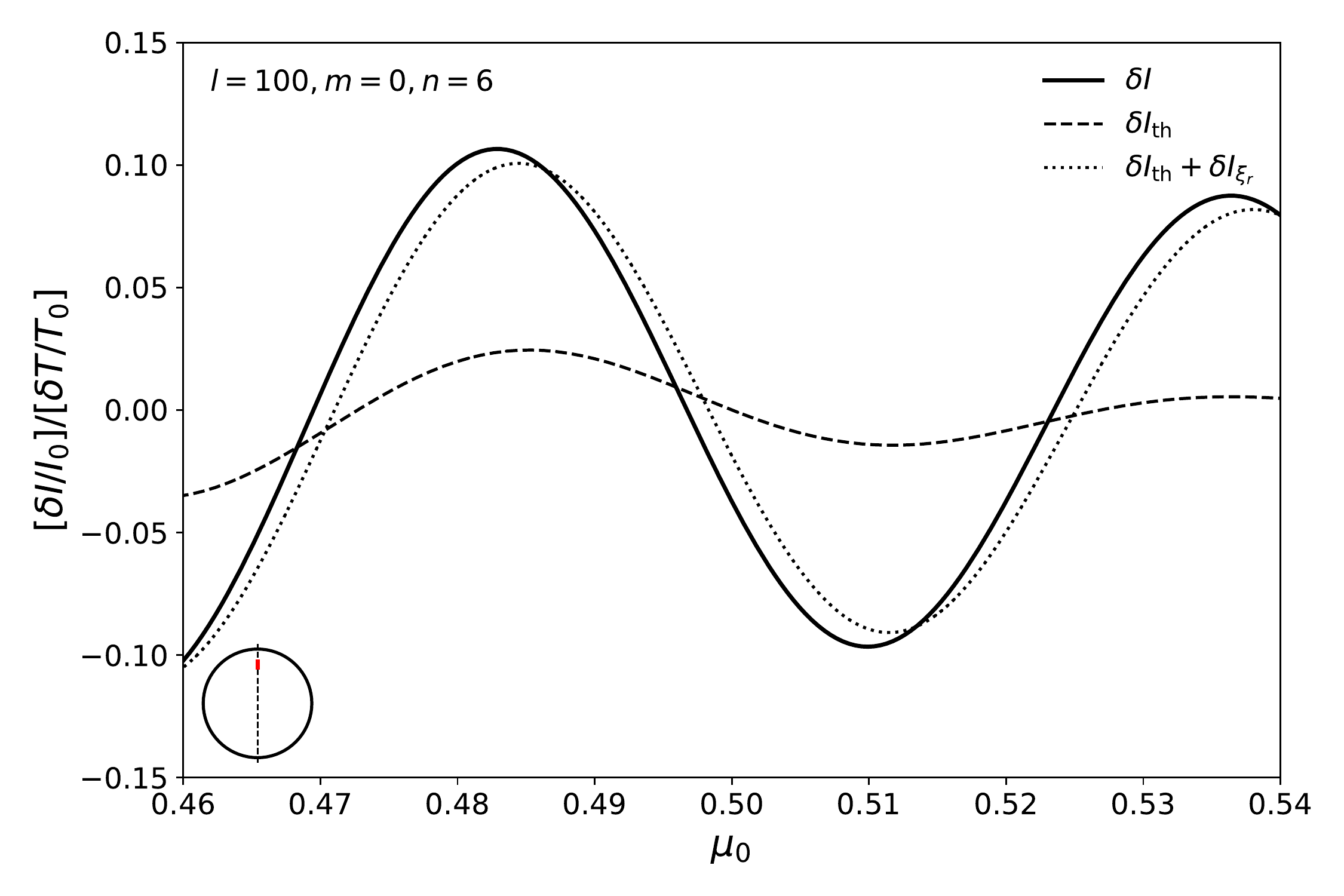}
    \includegraphics[width=0.49\linewidth]{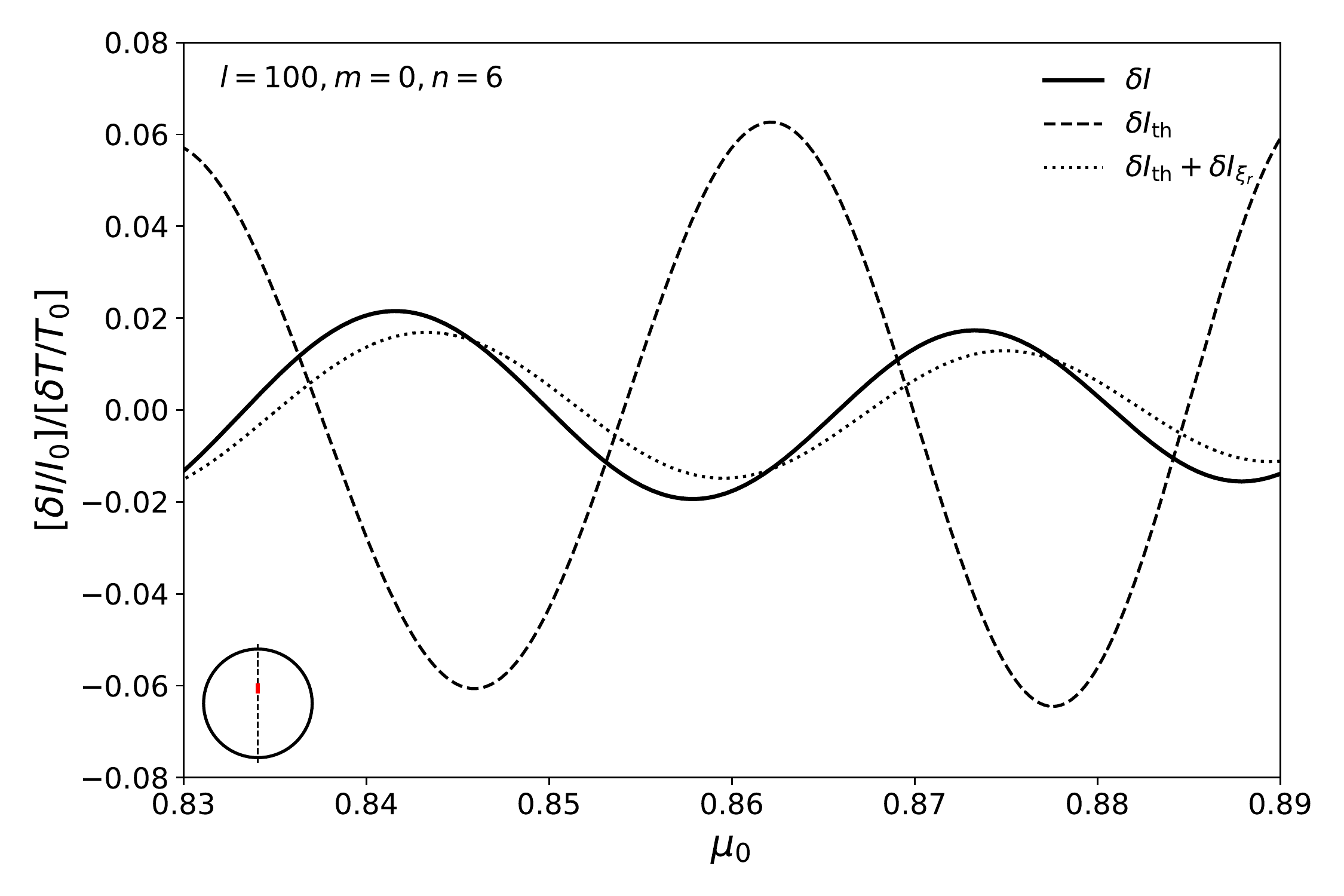}\\ 
    \includegraphics[width=0.49\linewidth]{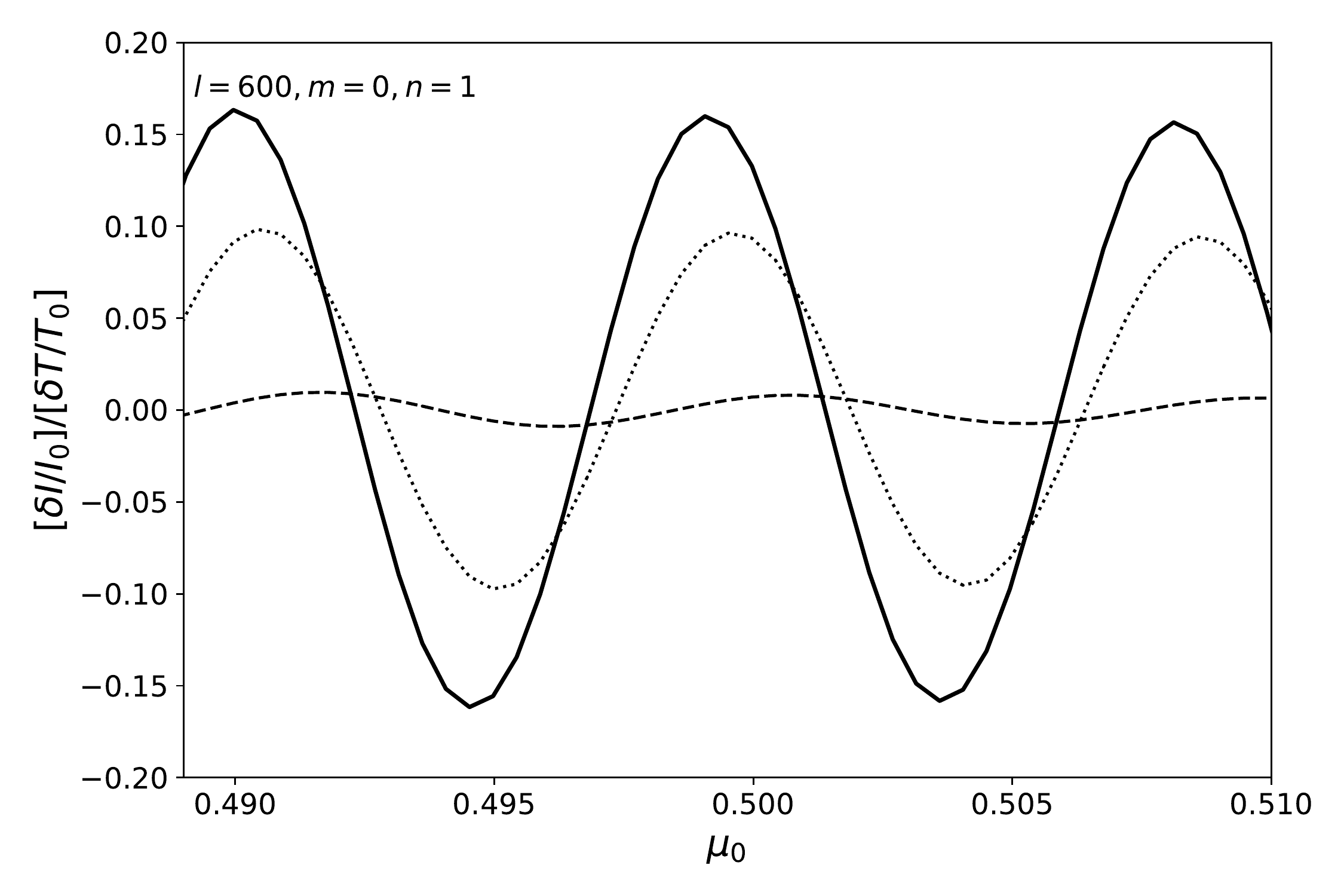}
    \includegraphics[width=0.49\linewidth]{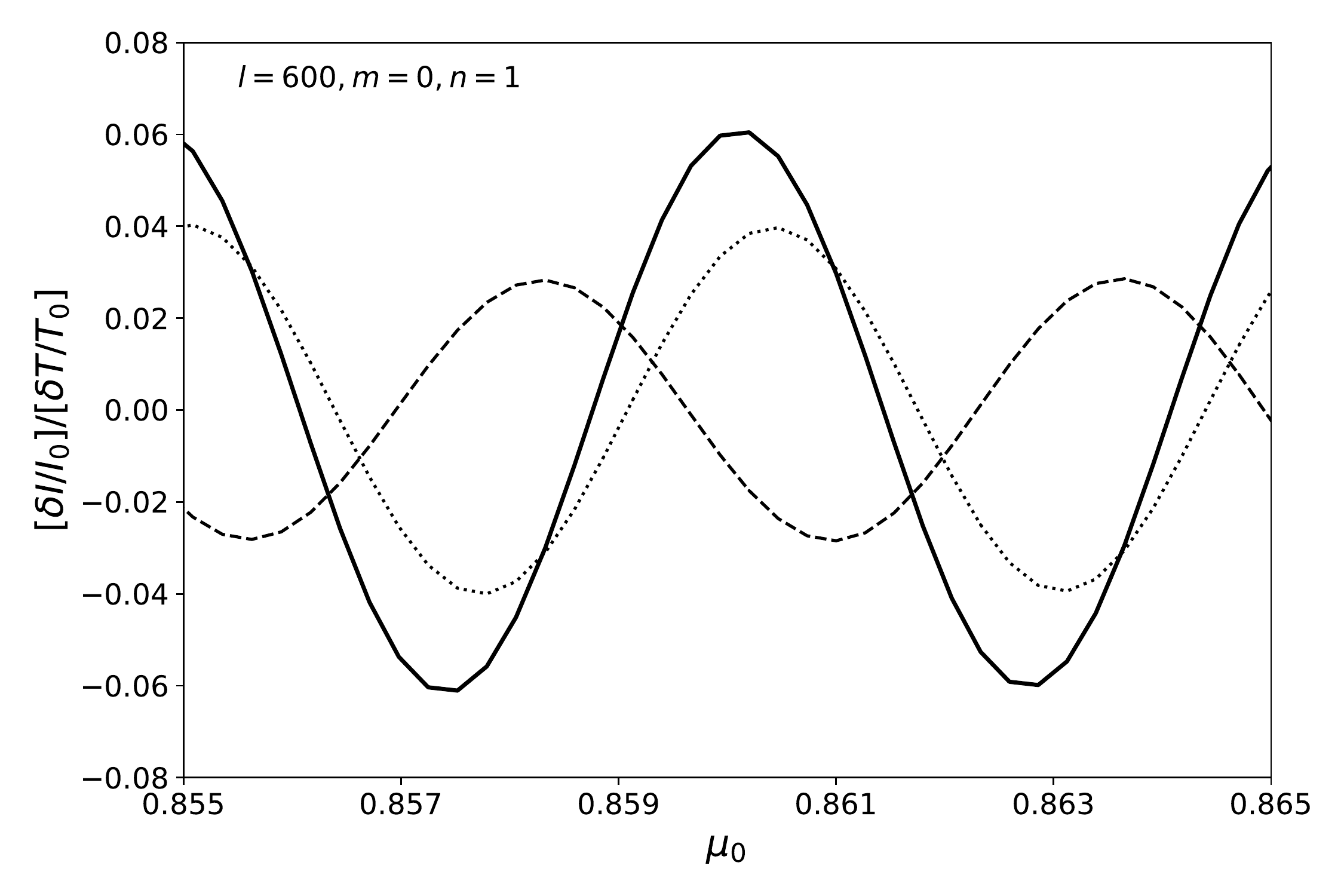}\\
   \caption{Normalised intensity perturbations caused by $l=100$ (top) and $l=600$ (bottom) modes at frequencies around 3~mHz at different center-to-limb distances: $\mu_0$ around  $0.85$ (left panels) and around $0.5$ (right panels). The observer frame showing where intensity perturbations are calculated is shown at the bottom left corners of the upper panels. 
   }
   \label{fig:Intensity_pert_500nm_highdegree}
\end{figure*}

In Sect.~\ref{sect:low}, we showed the importance of the geometrical term to compute the emergent intensity for a radial mode. Here, we analyse this effect for the modes where an extra contribution comes from the horizontal displacement. We select two modes, one at $l=100$ and one at $l=600$ for which both radial and horizontal displacements are significant (see the ratio of the horizontal-to-radial displacements in Table~\ref{tab:modes}). Figure~\ref{fig:Intensity_pert_500nm_highdegree} shows the normalised intensity perturbations with different contributors included (i.e. $\delta_I$, $\delta I_{\rm th} + \delta I_{\xi_r}$, and $\delta I_{\rm th}$) for different center-to-limb distances centred around $\theta = 30^{\rm o}$ and $\theta = 60^{\rm o}$  (corresponding to $\mu_0 = 0.86$ and $\mu_0=0.5$, respectively). 

Like for the radial mode, the differences between $\delta I$ and $\delta I_{\rm th}$ are important in terms of phase and amplitude, thus the thermodynamical quantities are not sufficient to describe accurately the emergent intensity.

Moreover, we compare $\delta I$ to $\delta I_{\rm th} + \delta I_{\xi_r}$ to see the importance of the horizontal displacement $\xi_\theta$ which was not taken into account in previous studies. As $\xi_r$ and $\xi_\theta$ are not in phase, it creates an additional phase shift and modifies the amplitude of the observed emergent intensity. The differences increase towards the limb but the importance of the horizontal term is already visible at a latitude of $30^\circ$ ($\mu = 0.86$). From those examples, we can ascertain that the higher the degree of the mode, the higher the differences between $\delta_I$ and $\delta I_{\rm th} + \delta I_{\xi_r}$.
 
\section{Summary and discussion} 
\label{summary}

We provided a detailed computation of the relationship between the eigenfunctions of solar p modes and continuum intensity, which is one of the helioseismic observables. We derived an  analytic expression for the emergent intensity perturbations caused by p-mode oscillations. These oscillations perturb the thermodynamical quantities in the solar atmosphere, as well as the integration path across the atmospheric layers and the position on the disk from which the radiation reaches the observer.  

The thermodynamical component contains only the perturbations in temperature and pressure in the atmosphere caused by the oscillations. It leads to perturbations of the source function and opacity in the radiative transfer equation. For the opacity, we considered absorption coefficients caused by bound-free and free-free transitions of different species, metal photo ionization, Rayleigh scattering on HI and HeI, and Thomson scattering on free electrons. We showed that to compute accurately emergent intensity all sources of opacity should be taken into account, and not only the bound-free transition of $\rm{H^-}$ which is the main contributor to the continuum opacity in the visible spectral range.

The geometrical component includes the perturbation to the geometrical ray path (including its direction) and the resulting perturbation to the optical depth. In doing so, we took into account both the radial and the horizontal displacements of the modes. The horizontal displacement has a negligible effect for the low degree modes around $3~\rm{mHz}$, however it cannot be neglected for low-frequency modes and high-degree modes for which the ratio between horizontal and radial displacements is significant. This may lead to  amplitude changes and phase shifts between the temperature and intensity perturbations, which increase towards the limb. 

The computation of the continuum intensity at one wavelength takes only 3~s for 2000 points along center-to-limb distance, which is not an obstacle for global and local helioseismology applications.
We presented computations in the continuum at 500~nm, however there is no limitation on the choice of wavelength.  In future work, we will study the emergent intensity perturbations along a spectral line, in order to synthesis  Doppler velocity observations.

\begin{acknowledgement}
We are thankful to Veronika Witzke for providing us with the MPS-ATLAS code for opacity calculation, Vincent B\"oning for providing us with the ADIPLS eigenfunctions, as well as Aaron Birch, Jesper Schou and Alexander Shapiro for useful discussions. This work was supported in part by a Max Planck Society grant ``Preparations for PLATO Science'' and German Aerospace Center (DLR) grants ``PLATO Data Center'' $50$OO$1501$ and $50$OP$1902$. LG and DF acknowledge partial support from ERC Synergy Grant WHOLE~SUN 810218 and Deutsche Forschungsgemeinschaft (DFG, German Research Foundation) through SFB~1456/432680300 Mathematics of Experiment, project C04.
\end{acknowledgement}

\bibliographystyle{aa}
\bibliography{References}
\clearpage
\onecolumn
\begin{appendix} \section{Derivation of $\delta \mu$} \label{Append}

In this appendix, we derive the analytic expression for the perturbation of the angle of incidence $\delta\mu = \mu - \mu_0$.

The position vector of a point in the perturbed model is  
\begin{equation}
    \boldsymbol{r} = \boldsymbol{r}_0 + \boldsymbol{\xi} = (r_0+\xi_r) \ \be_r+\xi_\theta \ \be_\theta + \xi_\phi \ \be_\phi,
\end{equation}
\noindent where $\boldsymbol{r}_0$ is the unperturbed position vector and $r_0$ is the solar radius in the background model. 
The perturbation $\delta \mu$ is then given by 
\begin{equation}
\delta \mu  = (\hat{\boldsymbol{n}} - \hat{\boldsymbol{n}}_0) \cdot \be_{\textrm{obs}},   
\label{Eq:delta_mu_def}
\end{equation}
\noindent 
where $\hat{\boldsymbol{n}}_0 = \be_r$ and $\hat{\boldsymbol{n}} = {\boldsymbol{n}}/{\Vert \boldsymbol{n}\Vert}$ where  $\boldsymbol{n}$ is the normal to the perturbed surface element defined as
\begin{equation}
	\boldsymbol{n} = \frac{\partial \boldsymbol{r}}{\partial \theta} \times \frac{\partial \boldsymbol{r}}{\partial \phi} = \left(\frac{\partial r_\theta}{\partial \theta} \frac{\partial r_\phi}{\partial \phi}
    - \frac{\partial r_\phi}{\partial \theta} \frac{\partial r_\theta}{\partial \phi}\right) \be_r
    + \left(\frac{\partial r_\phi}{\partial \theta} \frac{\partial r_r}{\partial \phi}
    - \frac{\partial r_r}{\partial \theta} \frac{\partial r_\phi}{\partial \phi}\right)  \be_\theta
    + \left(\frac{\partial r_r}{\partial \theta} \frac{\partial r_\theta}{\partial \phi}
    - \frac{\partial r_\theta}{\partial \theta} \frac{\partial r_r}{\partial \phi}\right) \be_{\phi}. 
    \label{Eq:vect_prod}
\end{equation}

\noindent  The term ${\partial \boldsymbol{r}}/{\partial \theta}$ can be computed as follows
\begin{align}
\frac{\partial \boldsymbol{r}}{\partial \theta} &= \frac{\partial} {\partial \theta} \left[ (r_0 + \xi_r) \ \be_r \right] + \frac{\partial} {\partial \theta} ( \xi_\theta \ \be_\theta ) + 
    \frac{\partial} {\partial \theta} ( \xi_\phi \ \be_\phi ) \nonumber \\
    &= 
    (r_0 + \xi_r) \ \frac{\partial \be_r}{\partial  \theta} +  \frac{\partial \xi_r}{\partial \theta} \ \be_r + \xi_\theta \ \frac{\partial \be_\theta}{\partial \theta} +   \frac{\partial \xi_\theta}{\partial \theta} \ \be_\theta + \xi_\phi \  \frac{\partial \be_\phi}{\partial \theta} +   \frac{\partial \xi_\phi}{\partial \theta} \ \be_\phi \nonumber \\
    &= 
    \left(\frac{\partial \xi_r}{\partial \theta} - \xi_\theta\right) \ \be_r +  \left(r_0 + \xi_r+\frac{\partial \xi_\theta}{\partial \theta}\right) \ \be_\theta + \frac{\partial \xi_\phi}{\partial \theta} \ \be_\phi, 
  \label{Eq:der_theta}
\end{align}
where we have used the derivative of the unit vectors in spherical coordinates
\begin{equation}
 \frac{\partial \be_r}{\partial \theta} = \be_{\theta},  \quad
 \frac{\partial \be_\theta}{\partial \theta} = -\be_r, \quad
 \frac{\partial \be_\phi}{\partial \theta} = 0.
 \end{equation}
Similarly, we can derive the expression for $\frac{\partial \boldsymbol{r}}{\partial \phi}$
\begin{align}
    \frac{\partial \boldsymbol{r}}{\partial \phi} 
    &= 
    \frac{\partial} {\partial \phi} \left[ (r_0 + \xi_r) \ \be_r \right] + \frac{\partial} {\partial \phi} ( \xi_\theta \ \be_\theta ) + 
    \frac{\partial} {\partial \phi} ( \xi_\phi \ \be_\phi ) \nonumber \\
    &= 
    (r_0 + \xi_r)\frac{\partial \be_r}{\partial  \phi} + \be_r \frac{\partial \xi_r}{\partial \phi} + \xi_\theta  \frac{\partial \be_\theta}{\partial \phi} + \be_\theta  \frac{\partial \xi_\theta}{\partial \phi} + \xi_\phi  \frac{\partial \be_\phi}{\partial \phi} + \be_\phi  \frac{\partial \xi_\phi}{\partial \phi} \nonumber \\
    &=
\left(\frac{\partial \xi_r}{\partial \phi} - \xi_\phi \sin \theta\right) \ \be_r + \left(\frac{\partial \xi_\theta}{\partial \phi} - \xi_\phi \cos \theta\right) \ \be_\theta 
    + \left((r_0 +\xi_r) \sin \theta + \xi_\theta \cos\theta +\frac{\partial \xi_\phi}{\partial \phi}\right) \ \be_\phi ,
    \label{Eq:der_phi}
\end{align}
where we used that
\begin{equation}
     \frac{\partial \be_r}{\partial \phi} = \sin \theta \ \be_\phi,  \quad
 \frac{\partial \be_\theta}{\partial \phi} = \cos \theta \ \be_\phi, \quad
 \frac{\partial \be_\phi}{\partial \phi} = -\cos \theta \ \be_\theta-\sin\theta \ \be_r.
\end{equation}

\noindent Using Eq.~(\ref{Eq:der_theta}), Eq.~(\ref{Eq:der_phi}), and Eq.~(\ref{Eq:vect_prod}) we obtain
\begin{align}
 \boldsymbol{n} 
 = &   \left(r_0+\xi_r+\frac{\partial \xi_\theta}{\partial \theta} \right) \left( (r_0 +\xi_r)\sin \theta + \xi_\theta \cos \theta + \frac{\partial \xi_\phi}{\partial \phi} \right)\ \be_r \nonumber \\  
 &+ \left[\frac{\partial \xi_\phi}{\partial \theta} \left( \frac{\partial \xi_r}{\partial \phi} - \xi_\phi \sin \theta \right) - \left(\frac{\partial \xi_r}{\partial \theta}-\xi_\theta \right) \left( (r_0+\xi_r)\sin\theta +\xi_\theta \cos \theta +\frac{\partial \xi_\phi}{\partial \phi} \right) \right] \ \be_\theta \nonumber \\   
 &+ \left[ \left( \frac{\partial \xi_r}{\partial \theta}-\xi_\theta \right) \left( \frac{\partial \xi_\theta}{\partial \phi}-\xi_\phi \cos\theta \right) - \left(r_0+\xi_r+\frac{\partial \xi_\theta}{\partial \theta}\right) \left(\frac{\partial \xi_r}{\partial \phi} - \xi_\phi \sin\theta \right)\right] \ \be_\phi. 
 \label{Eq:vect_prod_new}
\end{align}
Neglecting the second-order terms in Eq.~(\ref{Eq:vect_prod_new}) the expression for the normal vector becomes
\begin{equation}
 \boldsymbol{n} = 
 \left(r_0^2 \sin \theta + 2r_0\xi_r\sin\theta+r_0 \xi_\theta \cos\theta +r_0 \frac{\partial \xi_\phi}{\partial \phi} +r_0\sin\theta \frac{\partial \xi_\theta}{\partial \theta}\right) \ \be_r
 +\left(-r_0\sin\theta \frac{\partial \xi_r}{\partial \theta}+r_0\xi_\theta\sin\theta\right)  \ \be_\theta
 + \left(r_0\xi_\phi\sin\theta - r_0\frac{\partial \xi_r}{\partial \phi}\right) \ \be_\phi. 
\end{equation}
We normalize the normal vector and take only the first-order terms into account:
\begin{equation}
    \hat {\boldsymbol{n}} = \frac{\boldsymbol{n}}{\Vert \boldsymbol{n} \Vert} =\be_r + \left(\frac{\xi_\theta}{r_0}- \frac{1}{r_0}\frac{\partial \xi_r}{\partial \theta}\right) \ \be_\theta + \left(\frac{\xi_\phi}{r_0} - \frac{1}{r_0\sin\theta}\frac{\partial \xi_r}{\partial \phi}\right) \ \be_\phi.
\end{equation}

\noindent It follows that
\begin{equation}
    \delta \mu = (\hat{\boldsymbol{n}} - \hat{\boldsymbol{n}}_0)\cdot\be_{\textrm{obs}} = \left(\frac{\sin{i}\cos{\theta}\cos{\phi} - \cos{i}\sin{\theta}}{r_0}\right)\left(\xi_\theta - \frac{\partial \xi_r}{\partial \theta}\right) - \frac{\sin{i} \sin{\phi}}{r_0\sin{\theta}} \left(\xi_\phi \sin{\theta} - \frac{\partial \xi_r}{\partial \phi} \right),
   \label{Eq:delta_mu}
\end{equation}
where we used 
\begin{equation}
    \be_{\textrm{obs}} = \left( \sin{i} \sin{\theta} \cos{\phi} + \cos{i}\cos{\theta}\right) \ \be_r 
    + \left( \sin{i} \cos{\theta} \cos{\phi} - \cos{i} \sin{\theta} \right) \ \be_\theta 
    - \sin{i} \sin{\phi} \ \be_\phi. \label{eq:eObsSph}
\end{equation}

This expression for $\delta \mu$ can be evaluated for  special cases found in the literature. For example, choosing $i = 90^\circ$ and $\phi = 0$, we have $\mu_0 = \sin{\theta}$ and
 \begin{equation}
   \delta \mu =  \frac{\cos{\theta}}{r_0}\left(\xi_\theta-\frac{\partial\xi_r}{\partial \theta}\right) = 
    \frac{1-\mu_0^2}{r_0}\left(\frac{\xi_\theta}{\sqrt{1-\mu_0^2}}-\frac{\partial \xi_r}{\partial \mu_0}\right),
    \label{Eq:delta_mu_Heind}
\end{equation}
which is the expression derived by \citet{1994AAS..105..447H}. If we further assume  $\xi_\theta \ll \xi_r$, then $\delta \mu$ becomes  
\begin{equation}
	\delta\mu = \frac{\mu_0^2 - 1}{r_0}\frac{\partial\xi_r}{\partial \mu_0},
\end{equation}
as in \citetalias{1993AA...268..309T}. 

 \end{appendix}

\end{document}